\documentclass[10pt,twocolumn,letterpaper]{article}

\usepackage{iccv}
\usepackage{times}
\usepackage{epsfig}
\usepackage{graphicx}
\usepackage{amsmath}
\usepackage{amssymb}
\usepackage{subcaption}
\usepackage{adjustbox}
\usepackage{booktabs}
\usepackage{authblk}
\graphicspath{{./trans_fig/}}
\usepackage[breaklinks=true,bookmarks=false]{hyperref}

\iccvfinalcopy % *** Uncomment this line for the final submission

\def\iccvPaperID{38} % *** Enter the ICCV Paper ID here

\ificcvfinal\fi

\begin{document}

%%%%%%%%% TITLE
\title{DISGAN: Wavelet-informed Discriminator Guides GAN to MRI Super-resolution with Noise Cleaning}

\author[1,*]{Qi Wang}
\author[1]{Lucas Mahler}
\author[2,1]{Juliu Steiglechner}
\author[2,1]{Florian Birk}
\author[2,1]{Klaus Scheffler}
\author[2,1]{Gabriele Lohmann}
\affil[1]{Max Planck Institute for Biological Cybernetics, Tübingen, Germany}
\affil[2]{University Hospital Tübingen, Tübingen, Germany}
\affil[*]{corresponding author: {\tt\small qi.wang@tuebingen.mpg.de}}
%\author{Qi Wang\\
%Max Planck Institute for Biological Cybernetics\\
%Tübingen, Germany\\
%{\tt\small qi.wang@tuebingen.mpg.de}
%% omit the following lines up until the closing ``}''.
%% Additional authors and addresses can be added with ``\and'',
%% just like the second author.
%% To save space, use either the email address or home page, not both
%\and
%Lucas Mahler\\
%Max Planck Institute for Biological Cybernetics\\
%Tübingen, Germany\\
%{\tt\small lucas.mahler@tuebingen.mpg.de}
%\and
%Julius Steiglechner\\
%University Hospital of Tübingen\\
%Tübingen, Germany\\
%{\tt\small julius.Steiglechner@tuebingen.mpg.de}
%\and
%Florian Birk\\
%University Hospital of Tübingen\\
%Tübingen, Germany\\
%{\tt\small florian.brik@tuebingen.mpg.de}
%\and
%Klaus Scheffler\\
%University Hospital of Tübingen\\
%Tübingen, Germany\\
%{\tt\small klaus.scheffler@tuebingen.mpg.de}
%\and
%Gabriele Lohmann\\
%University Hospital of Tübingen\\
%Tübingen, Germany\\
%{\tt\small lohmann@tuebingen.mpg.de}
%}

\maketitle
\ificcvfinal\thispagestyle{empty}\fi

%%%%%%%%% ABSTRACT
\begin{abstract}
MRI super-resolution (SR) and denoising tasks are fundamental challenges in the field of deep learning, which have traditionally been treated as distinct tasks with separate paired training data. In this paper, we propose an innovative method that addresses both tasks simultaneously using a single deep learning model, eliminating the need for explicitly paired noisy and clean images during training. Our proposed model is primarily trained for SR, but also exhibits remarkable noise-cleaning capabilities in the super-resolved images.
Instead of conventional approaches that introduce frequency-related operations into the generative process, our novel approach involves the use of a GAN model guided by a frequency-informed discriminator. To achieve this, we harness the power of the 3D Discrete Wavelet Transform (DWT) operation as a frequency constraint within the GAN framework for the SR task on magnetic resonance imaging (MRI) data.
Specifically, our contributions include: 1) a 3D generator based on residual-in-residual connected blocks; 2) the integration of the 3D DWT with $1\times 1$ convolution into a DWT+conv unit within a 3D Unet for the discriminator; 3) the use of the trained model for high-quality image SR, accompanied by an intrinsic denoising process. We dub the model "Denoising Induced Super-resolution GAN (DISGAN)" due to its dual effects of SR image generation and simultaneous denoising.
Departing from the traditional approach of training SR and denoising tasks as separate models, our proposed DISGAN is trained only on the SR task, but also achieves exceptional performance in denoising. The model is trained on 3D MRI data from dozens of subjects from the Human Connectome Project (HCP) and further evaluated on previously unseen MRI data from subjects with brain tumours and epilepsy to assess its denoising and SR performance. Our code is available at \url{https://github.com/wqlevi/DISGAN}.
\end{abstract}

%%%%%%%%% BODY TEXT
\section{Introduction}
High-resolution MR images play a crucial role in providing detailed anatomical information essential for downstream MRI analysis. However, obtaining high-resolution (HR) MRI scans is a labor-intensive process prone to motion artifacts, particularly challenging for subjects with brain diseases such as brain tumors or epilepsy. Additionally, inherent physical limitations of MRI scanners and hardware introduce various types of noise into diagnostic or analytical images, leading to lower spatial resolution and loss of anatomical details. As a result, researchers have turned to deep learning methods to enhance the quality of existing MRI images, focusing on both denoising and Single Image Super-Resolution (SISR) techniques.

Conventional SR aims to recover HR MR images from lower-resolution images of the same subject, while denoising tasks focus on removing common noise sources such as Gaussian noise and motion artifacts to obtain cleaner image content. Traditionally, these tasks require separate training and paired datasets in most deep learning methods.

Both SISR and denoising are inherently inverse problems, and in the context of medical imaging, the challenge is further compounded by the curse of dimensionality. Moreover, access to 3D MRI data is significantly limited compared to the abundance of 2D image datasets, making training an SISR model on 3D medical images more demanding. As real-world medical images, such as brain MRI, contain vital anatomical information in all three dimensions, the primary focus of this paper lies in tackling the task of SISR on 3D volumetric MRI.

In the 2D domain, various studies have been conducted to achieve high image fidelity in SISR tasks using GAN architectures \cite{dong@srcnn,wang2018esrgan,srgan@ledig,vdsr}. Recently, other models, such as score-matching models \cite{scorematching@song}, diffusion probability models \cite{DDPM}, and transformers \cite{vit}, have also been employed to produce SR results with advanced image quality, both in metric scores and visual fidelity. In the 3D MRI domain, some of these models have been re-implemented, with the majority of them utilizing GANs \cite{DCSRN,srmri,ArSSR,mri@gradient}, continually pushing the boundaries of state-of-the-art performance.

However, training GAN models is known for its instability and sensitivity to parameter changes, requiring careful architecture design, especially in high-dimensional spaces with deep architectures. Wasserstein variations of GAN~\cite{wgan@Arjovsky,wgangp@Ishaan} have proven effective and popular in training SR models in the 3D MRI domain. Nevertheless, we observed that results from WGAN (or WGANGP) variations, such as DCSRN~\cite{DCSRN}, often suffer from blurring, likely due to convergence issues or oscillating around suboptimal solutions~\cite{dirac_gan,clcgan}.

In this paper, we propose a novel approach that incorporates instance noise to guide the training process, utilizes relativistic GAN loss~\cite{ragan} to accelerate convergence, and introduces a 3D Discrete Wavelet Transform (DWT)-informed discriminator to lead the generator towards minimal noise generation. The proposed model demonstrates stable training dynamics and produces high-quality results while significantly improving noise cleaning on both simulated noisy data and real-world images without additional training, outperforming other existing models.

Our work addresses the limitations of traditional SR models that heavily rely on fixed datasets, lacking robustness when applied to noisy images. In contrast, our proposed model, DISGAN, achieves better than state-of-the-art results in recovering detailed brain MRI structures while effectively cleaning noise, thereby presenting high-fidelity image content. The key contributions of our work are:

\begin{itemize}
\item We propose a GAN framework for the SR task that outperforms existing methods in restoring detailed structures.
\item We demonstrate the robustness of our proposed model to simulated noisy images during the SR process, guided by a frequency-informed discriminator.
\item We show the effectiveness of noise cleaning in our model, enhancing the quality of real-world clinical data, such as images of patients with epilepsy or brain tumors.
\end{itemize}

In summary, this paper presents a novel DISGAN model that not only enhances the resolution of MRI images but also effectively cleans noise, addressing a critical need in the field of medical imaging and advancing the state-of-the-art in MRI image super-resolution and denoising.
% TODO:
% adjust the subtitle positions

\section{Related works}
\subsection{Wavelet transforms}
Wavelet transforms are widely used in generative models to explicitly introduce frequency bias to the models in tasks like SR, denoising, and domain transfer\cite{DDUNet,MB}. Some of these works use GANs for better image generation, mostly implementing DWT explicitly in the generator. Our model includes the DWT functions in the discriminator, acting as a measurement in frequency space between generated distribution and the real one.%
\begin{figure*}[t]
\begin{center}
\includegraphics[width=1\linewidth]{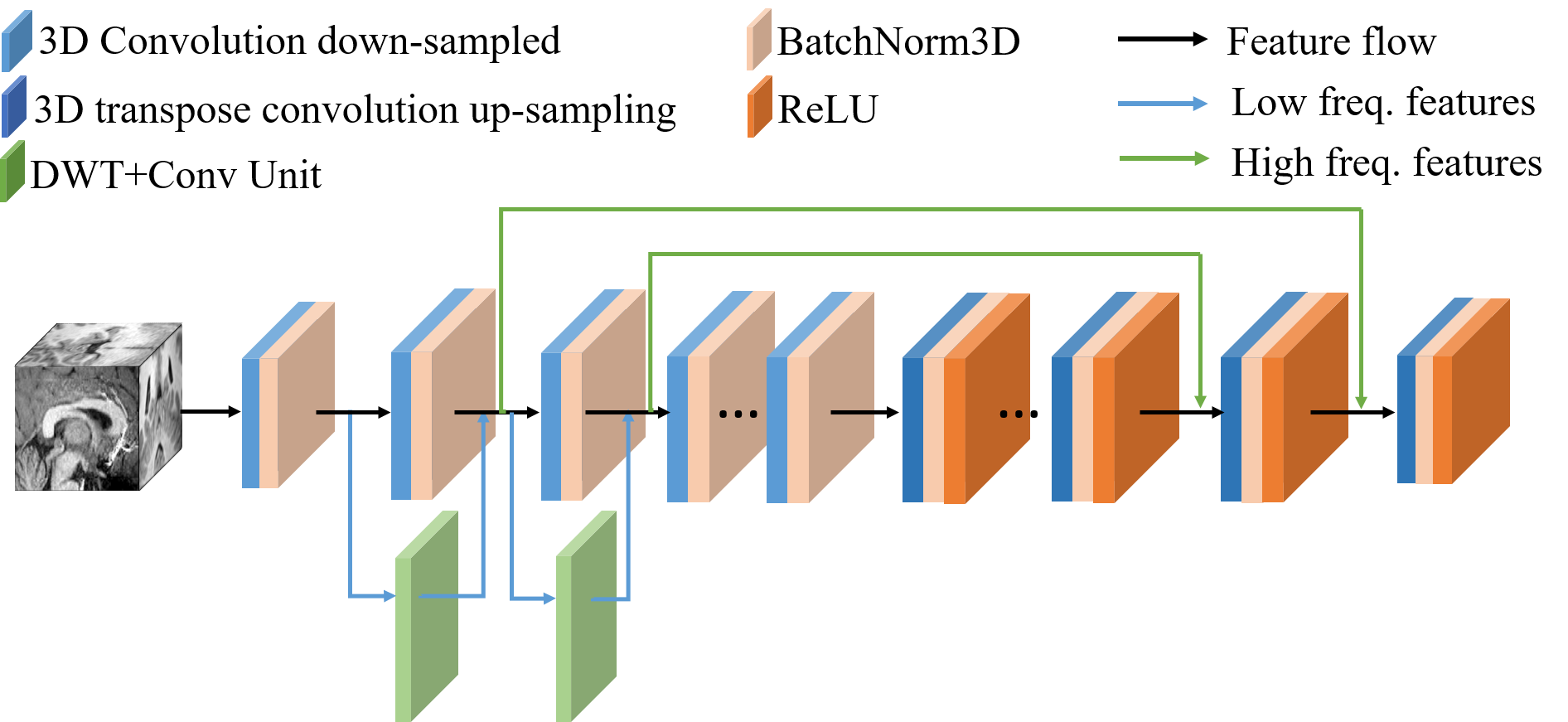} 
   \caption{The diagram of our DWT-informed discriminator. The input is HR or SR patch volumes, the output is a scoring tensor with the same spatial dimension as the input. The overall architecture follows a "U-shape" topology, where output after each blocks of the down-sampling and up-sampling branches are interconnected.}
\label{fig:whole_diagram}
\end{center}
\end{figure*}

\subsection{Super resolution in medical imaging}
Learning-based SISR has been extensively studied for both 2D and 3D images. In this task, neural networks aim to learn a non-linear mapping from low-resolution (LR) to high-resolution (HR) images by exploiting the powerful capabilities of convolutional neural networks (CNNs), as demonstrated in previous studies \cite{dong@srcnn,srgan@ledig,rcan,wang2018esrgan}.

In the 2D domain, SRCNN \cite{dong@srcnn} introduced the first end-to-end CNN architecture for LR to HR image mapping, significantly outperforming traditional SISR methods. Subsequent architectures such as SRGAN \cite{srgan@ledig}, VDSR \cite{vdsr}, RCAN \cite{rcan} and ESRGAN \cite{wang2018esrgan} further improved performance by incorporating GAN frameworks, perceptual loss, deeper architectures and attention mechanisms. More recent approaches using transformers \cite{vit, transformer_sr} and diffusion models \cite{scorematching@song, sr3} have achieved state-of-the-art quality in 2D SISR tasks. However, these methods typically require large amounts of training data.

In medical imaging, studies on 2D MR images have produced remarkable results, such as the squeeze-excitation attention network \cite{MRSR_SEN} and the transformer architecture \cite{transformer_mr}. However, 3D volumes of MR images are typically required for downstream analysis, making 3D models more favourable. Consequently, 3D models, such as those in \cite{srmri}, have demonstrated superiority over 2D models in SISR tasks for MR images.

This paper focuses only on 3D MR image training, and novel architectures have not yet been fully adapted to the 3D MRI domain due to its high dimensionality and limited data. As a result, GAN-based methods remain the mainstream choice for 3D MRI training due to their efficiency and high performance. Previous works, such as \cite{srmri, DCSRN, ArSSR, mri@gradient, ours_tmi,sr_uhf}, have explored GAN training, densely connected residual blocks, implicit neural representation and image gradients to achieve impressive 3D MR image SR results.

Despite the success of these studies, they are often constrained by having the same modality or image sequence as the training dataset, and the quality of the reconstruction can be limited in some cases.
\section{Methods}
\subsection{Network structure}
Here, we propose our DISGAN architecture, which consists of a generator, a discriminator as shown in Figure~\ref{fig:whole_diagram}, and a feature extractor. The generator consists of three Volumetric RRDB blocks~\cite{Huang_2017_CVPR} (VRRDB) and uses pixel shuffling as an upsampling operation. The feature extractor uses convolutional layers and activation layers in front of the multilayer perceptrons, which we train simultaneously with the GAN model. 

 The model takes paired LR and HR patches of the volumetric MRI and outputs an SR image.  Specifically, as for the generating part, the LR volume is the input to the generator to output the SR image. Then the feature extractor takes the pair of HR and SR to measure the feature-wise Euclidean distance. For the discriminator, the input is the pair of HR and SR images, and the output is a discrimination score. During training, we add linearly annealed Gaussian noise to stabilise the training. We use relativistic GAN loss as the target for adversarial training. 
\begin{figure}[!h]
   \includegraphics[width=\linewidth]{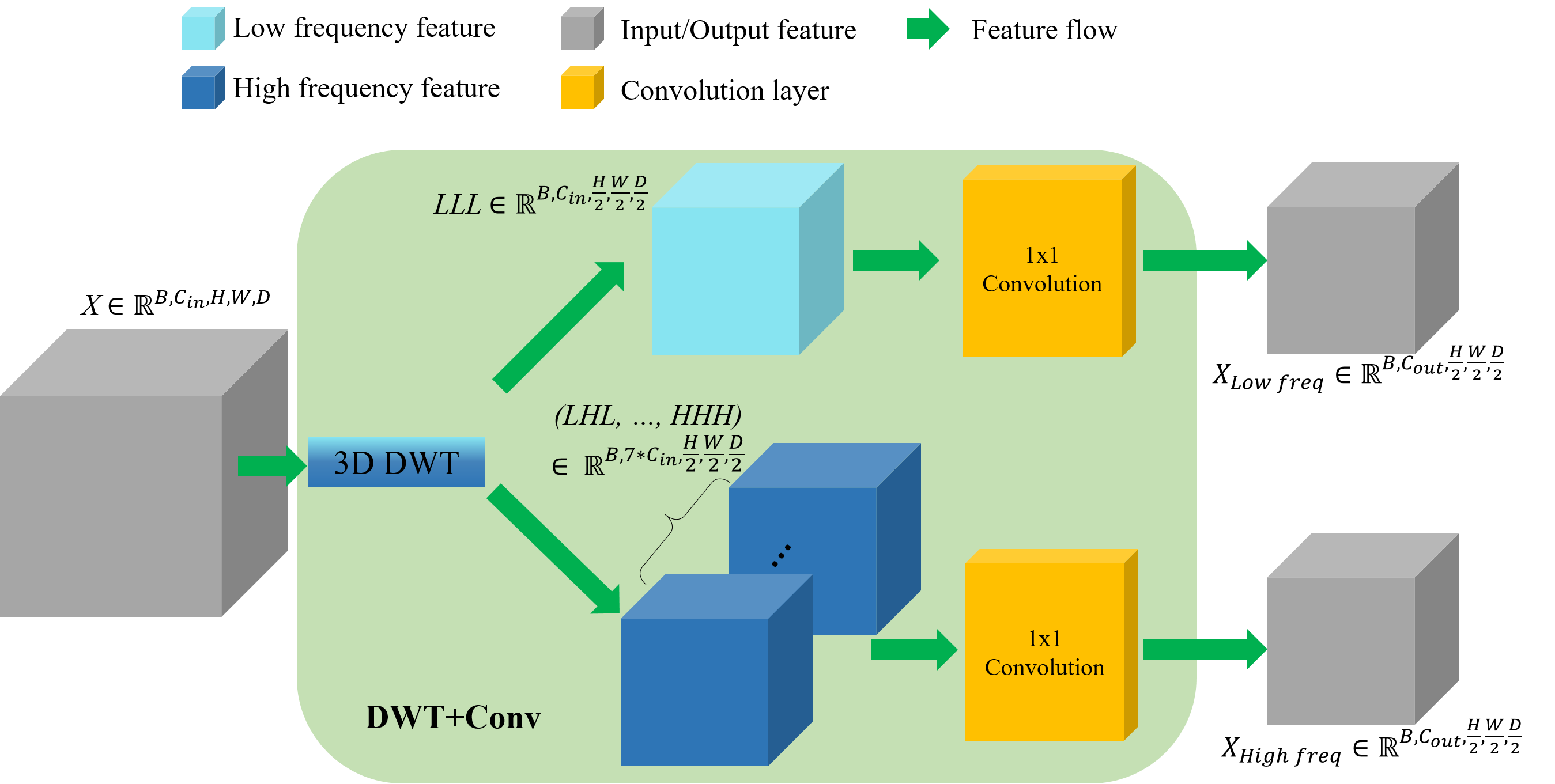}
   \caption{Diagram of the proposed DWT+conv units. Each DWT+conv unit takes high-resolution feature as input, and parse it through a 3D DWT using Haar wavelets while reducing the feature resolution by half. The low frequency feature is fed to a $1\times 1$ convolution layer to confine the numbers of output filters, whereas seven high frequency features are concatenated along filters' dimension before being fed into another $1\times 1$ convolution layer.}
   \label{fig:diagram}
\end{figure}
 \subsection{Wavelet informed discriminator}
We propose the Discrete Wavelet Transform Convolution Unit (DWT+conv) for extracting frequency-wise feature in the discriminator. As a fundamental operation of the DWT+conv, we use {\emph{3D Haar wavelet transform}} for calculating the sub-band frequency features, which are filtered into sub-bands comprising: one low frequency band ($LLL$), seven high frequency bands along different directions: $LHL, LLH, LHH, HLL, HHL, HLH, HHH$ (see Figure~\ref{fig:diagram}). After categorizing sub-frequency features, low frequency sub-band features are passed into a $1\times 1$ convolution to restrict the number of output filters; high frequency sub-band features are concatenated along the filters' dimension before being fed into another $1\times 1$ convolution for the same purpose.

\subsection{Objective function}
In order to generate images with both high perceptual quality and balanced training performance, we combine $\mathcal{L}_{1}$ loss in the image domain, perceptual loss~\cite{feifeili, srgan@ledig}, and relativistic GAN loss ($\mathcal{L}^{RaGAN}$)~\cite{ragan} as the overall objective function for the generator network. The parameters of the generator are constrained to learn both feature-wise ($\mathcal{L}^{perc}$) and image-wise ($\mathcal{L}^{pixel}$) representations of the ground truth, which the objective function seeks to minimize. The overall objective function of the generator is defined as:

\begin{equation}\label{eq:objfn}
  \mathcal{L}_{G} = \mathcal{L}^{perc} + \alpha \mathcal{L}^{pixel} + \beta \mathcal{L}_{G}^{RaGAN}
\end{equation}
where $\alpha = 0.01$ and $\beta = 0.005$ are empirical parameters for weighting image space $\mathcal{L}_{1}$ loss and relativistic averaged GAN loss for generator ($\mathcal{L}_{G}^{RaGAN}$). 

The discriminator and the feature extracting networks are updated based on their respective loss, $\mathcal{L}_{D}^{RaGAN}$ and $\mathcal{L}^{perc}$. 

\section{Experimental settings}
\subsection{Evaluation metrics}
Various metrics have been proposed to assess the accuracy of super-resolution results, such as peak signal-to-noise ratio (PSNR) and normalised-root-mean-square error (NRMSE). Since our goal is not only to maximise per-pixel similarity but also to correctly preserve anatomical structures, we use the Structural Similarity (SSIM) metrics to evaluate the quality of super-resolution results.

\subsection{Datasets}
To test our approach, we use several datasets from different scanners, imaging modalities or body parts. Three of these datasets are part of the Human Connectome Project~\cite{david2013hcp}\footnote{Data were provided [in part] by the Human Connectome Project, WU-Minn Consortium (Principal Investigators: David Van Essen and Kamil Ugurbil; 1U54MH091657) funded by the 16 NIH Institutes and Centers that support the NIH Blueprint for Neuroscience Research; and by the McDonnell Center for Systems Neuroscience at Washington University.}, the rest are from~\cite{knee}. We name the datasets according to the most distinctive attribute of each dataset, as described below.

As a pre-processing step, all images are standardised to have a mean of zero and a standard deviation of one. During training, each complete HR volume is patched into overlapping HR patches with a shape of $64\times 64\times 64$, with a step size of 16 along each dimension, the LR patches are simulated by linearly down-sampling the HR patches to $32\times 32\times 32$. All experiments use the model trained in this way on the Insample dataset.

\textbf{Dataset "Insample".} This dataset was downloaded from the \emph{Lifespan Pilot Project} as part of the Human Connectome Project (HCP)~\cite{david2013hcp}. It consists of T1-weighted (T1w) MR images of 27 healthy individuals aged 8 to 75 years (15 women) acquired on a Siemens 3T scanner. In our model, we divided them into 20 subjects for training and 7 for testing. The resolution of the ground truth images is $0.8mm$ isotropic, with a matrix size of $208\times 300\times 320$. 

\textbf{Dataset "Epilepsy".} This dataset was downloaded from one of the \emph{OpenNeuro} online repositories~\cite{epilepsy}. The repository we used contains mainly brain MRIs of epilepsy patients in T1w contrast. We use the image of one subject out of the 85 epilepsy subjects to evaluate the performance of the denoising SR. The image has a matrix size of $208\times 320\times 320$ with a resolution of $0.8 mm$.

\textbf{Dataset "Tumor".} This dataset is from the BraTS challenge validation dataset~\cite{BraTs}. The BraTS validation dataset contains several pathologically confirmed MRI scans, of which we use one from the brain tumour MRI with T1w. The image has a matrix size of $240\times 240\times 155$ with a resolution of $1 mm$ and has been skull-stripped in the dataset.
\begin{table}[h]
   \begin{center}
   \begin{tabular}{@{}l c c c c@{}} 
    \toprule
    Model            & PSNR $\uparrow$ & SSIM $\uparrow$ & NRMSE $\downarrow$\\ %[0.5ex] 
    \midrule
    Tri-linear       & 33.038 & 0.876 & 0.023 \\
    ESRGAN-3D~\cite{wang2018esrgan}        & 37.022 & 0.933 & 0.013 \\ 
    DCSRN~\cite{DCSRN}            & 37.635 & 0.954 & 0.013  \\
    ArSSR~\cite{ArSSR}	      & 28.038 & 0.280 & 0.048 \\
    Wang et al.~\cite{ours_tmi} & 36.922 & 0.943 & 0.013\\ %[1ex] 
    \textbf{DISGAN (ours)} & \textbf{39.342} & \textbf{0.962} & \textbf{0.006}\\ %[1ex] 
    \bottomrule
   \end{tabular}
   \caption{Quantitative comparison for SR models in dataset "Insample". Our model achieves the best performance in most of the metrics.}
   \label{table:insample}
\end{center}
\end{table}
\begin{figure*}[ht]
   \centering
	  \captionsetup[subfigure]{position=bottom}%
     \subcaptionbox{Tri-Linear}{\includegraphics[width=0.25\textwidth]{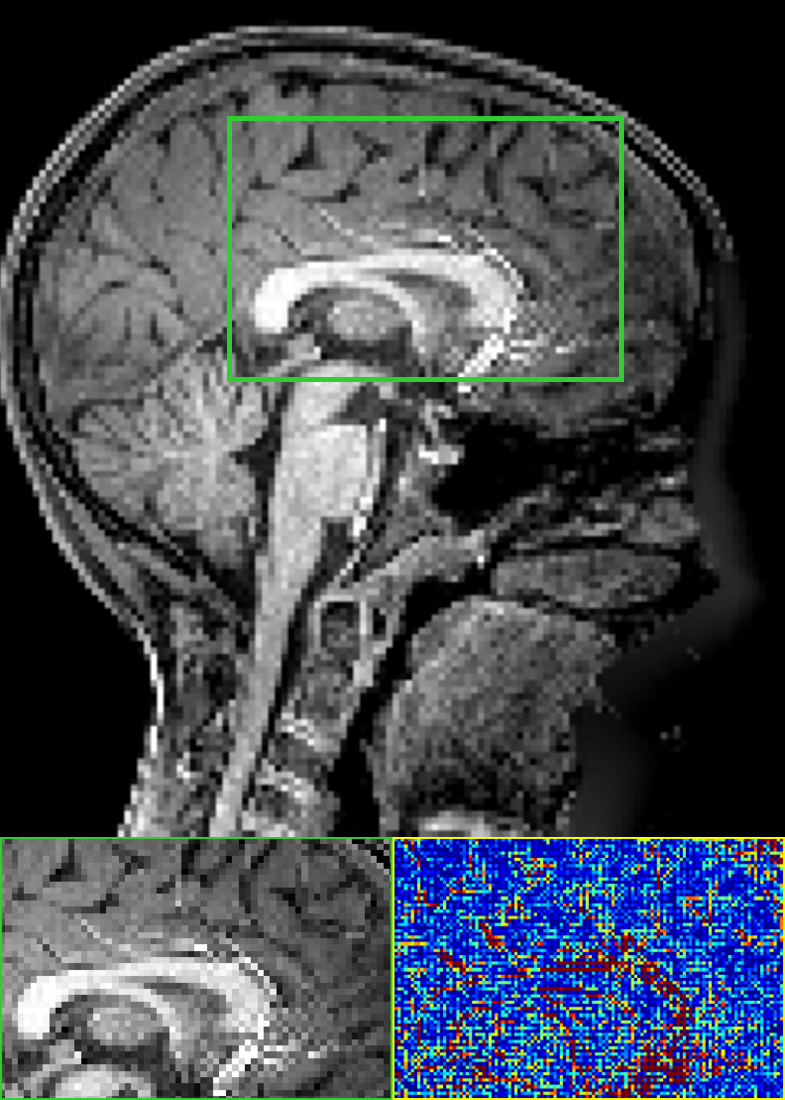}}
   \subcaptionbox{ESRGAN-3D\label{fig:short-a}}{\includegraphics[width=0.25\textwidth]{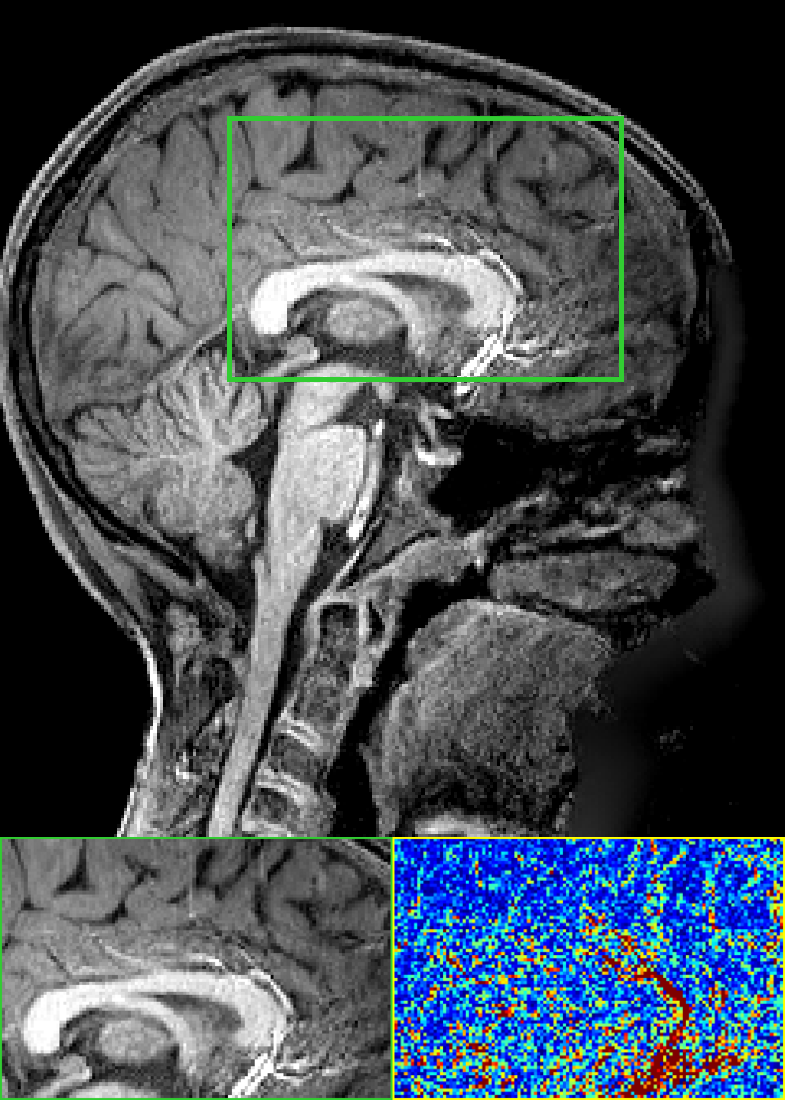}}
   \subcaptionbox{DCSRN}{\includegraphics[width=0.25\textwidth]{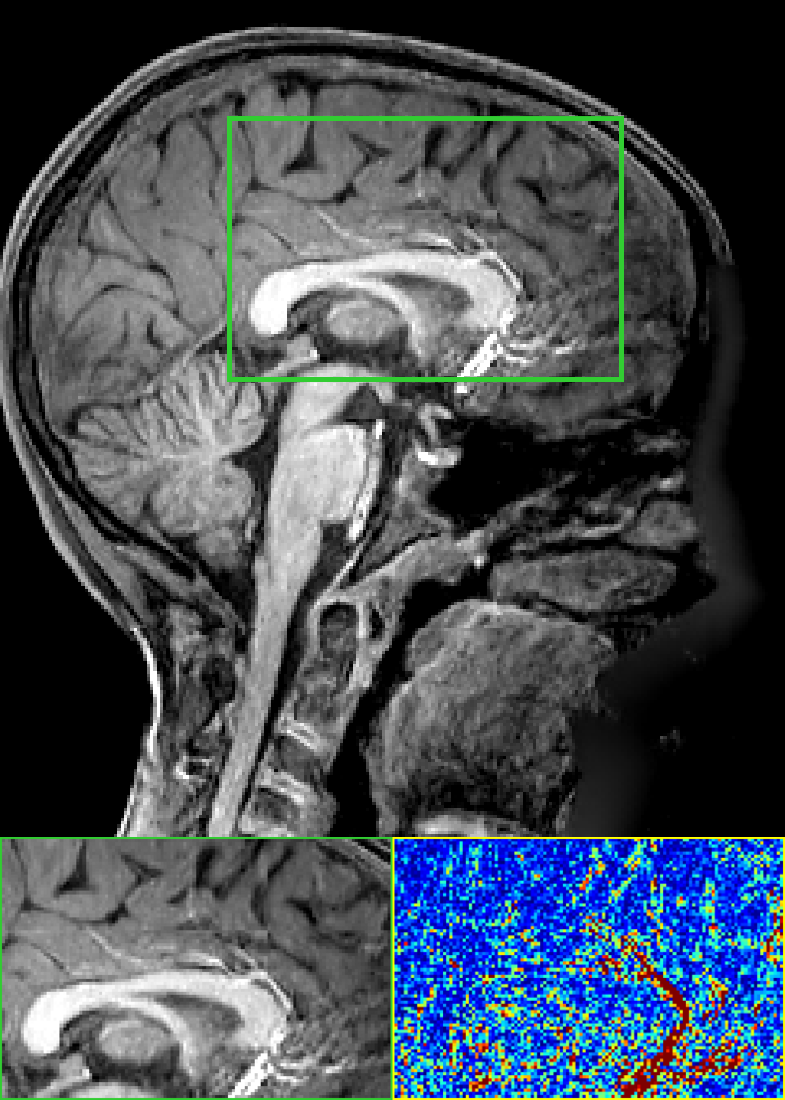}}
   \subcaptionbox{Wang et al.}{\includegraphics[width=0.25\textwidth]{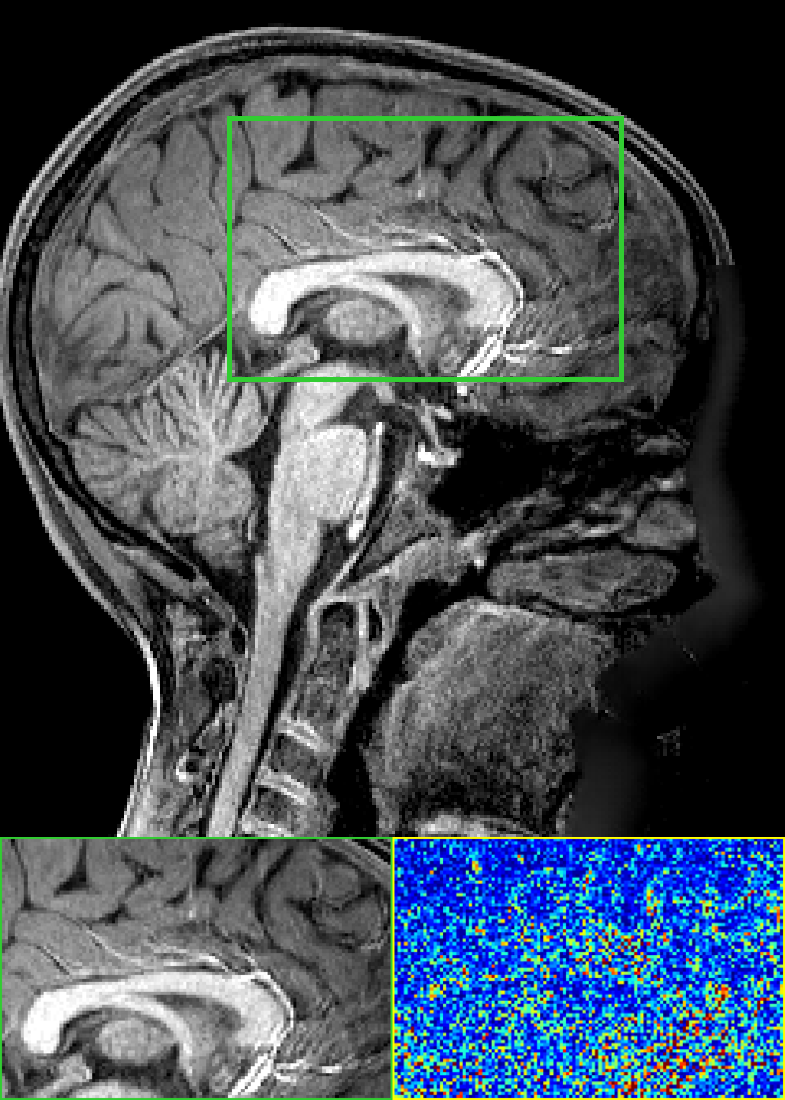}}
   \subcaptionbox{DISGAN (ours)}{\includegraphics[width=0.25\textwidth]{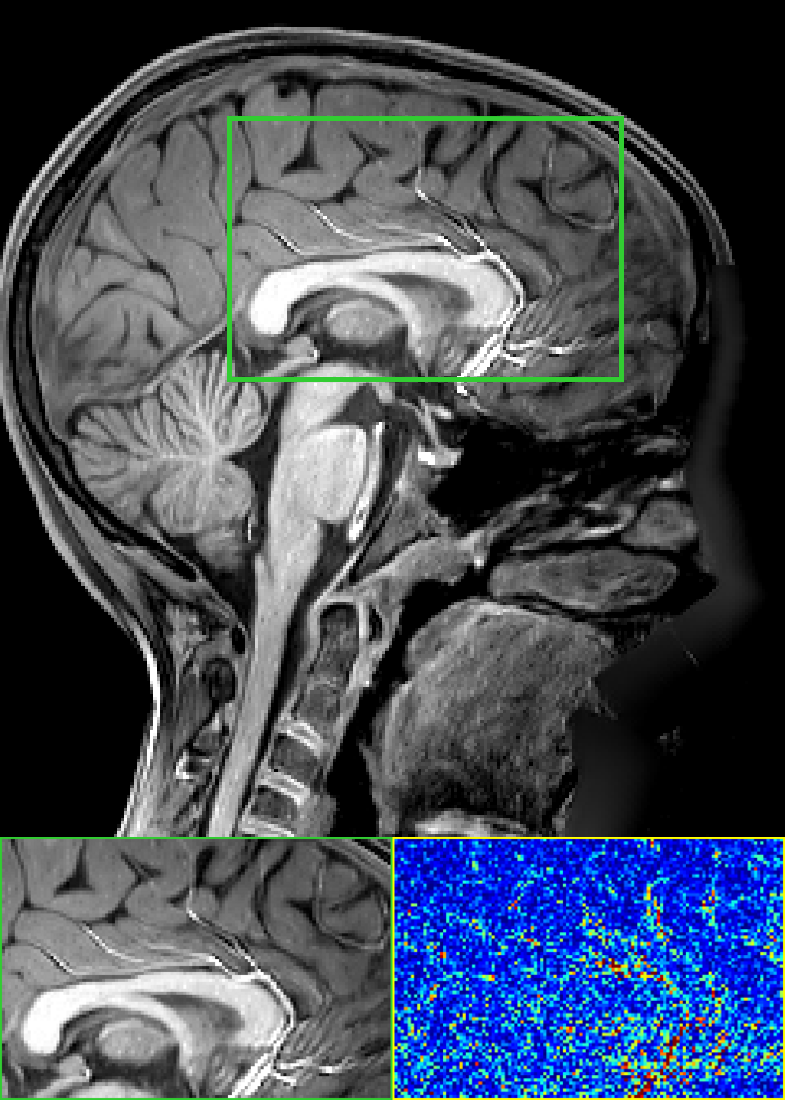}}
   \subcaptionbox{GT}{\includegraphics[width=0.25\textwidth]{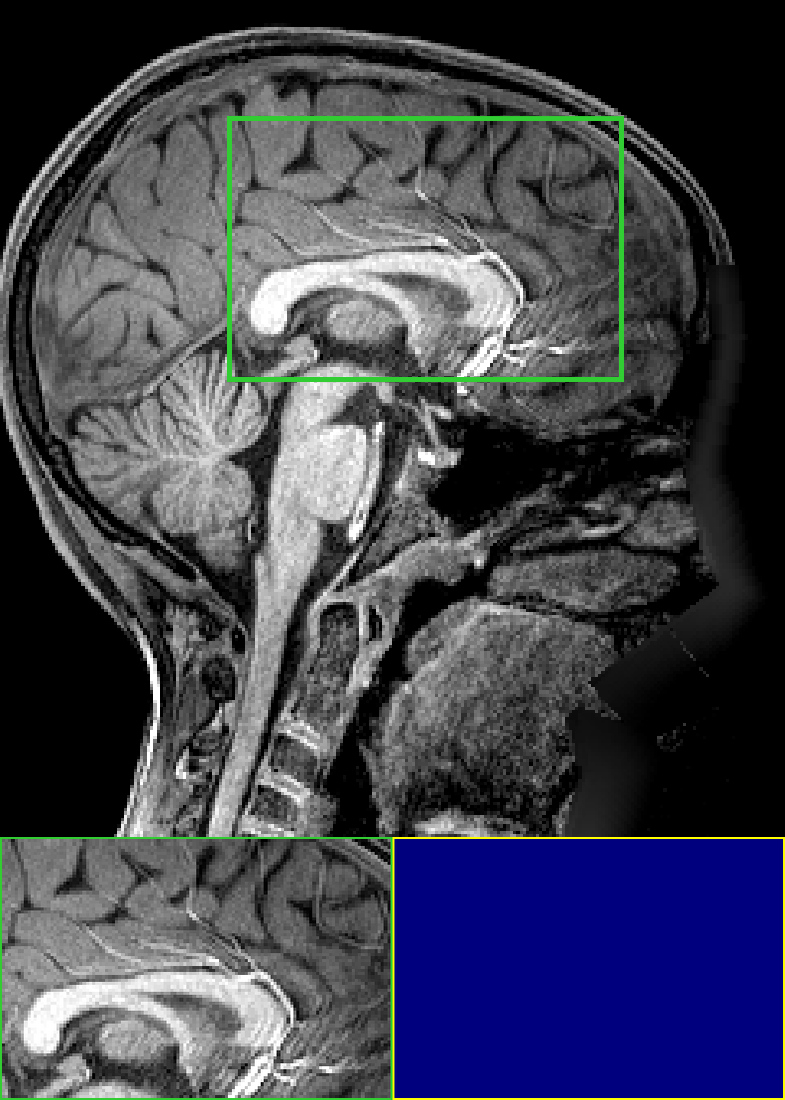}}%  
\caption{Super-resolution results of an exemplary subject using various methods. The zoom-in view and the colormaps (red for high and dark blue for low residual, using \textit{jet} colormap, same for all other colormaps) of its absolute residuals of the \textit{green boxes} are show below each column. Note that our method recovers detailed structures the most, such as the blood vessel (zoom in for better view). The residual maps highlight our model to have the best approximation, which is in line with the quantitative results in Table~\ref{table:insample}.}
\label{fig:insample}
\end{figure*}
\section{Results}
\subsection{Implementation details} \label{sec:implementation}
We trained a generator network, a critic network and a feature extractor network. The generator network consists of 3 residual-in-residual dense blocks\cite{wang2018esrgan}, which are densely connected residual units embedded without a batch norm layer. The critic network is the same as a discriminator without the final nonlinear activation layer. The feature extractor uses convolutional layers of ResNet10 before linear layers. All three networks are initialised by Kaiming initialisation\cite{kaiming_init} and optimised by Adam optimiser\cite{adam}, using coefficients of $\beta_{1} = 0.9$, $\beta_{2} = 0.999$ and a learning rate of $\gamma = 10^{-4}$. The variance of the instance noise decreases linearly in each iteration, from $\sigma = 1$ to $0$. Training is done simultaneously on the PyTorch framework\cite{PyTorch}, the generator and the discriminator network for 60000 iterations, on NVIDIA's Ampere 100 GPU.

%-------------------------------------------------------------------------
\subsection{High fidelity super-resolution on brain MRI}
In this experiment, we test the performance of our model against other 3D SR models on the \textit{"Insample"} dataset. For comparison, we tested our approach against four other approaches, namely ESRGAN-3D~\cite{wang2018esrgan}, DCSRN, and Wang et al.~\cite{ours_tmi}. ESRGAN was proposed in 2D, so we reconfigure it by replacing all 2D with 3D convolutions. Specifically, we pre-trained the first 54 layers of a 3D VGG-19 network on 3D MRI images from the same dataset. The DCSRN model is available in 3D, but we retrained the model on the "Insample" dataset.
\begin{figure*}[ht]
   \centering
   \captionsetup[subfigure]{position=bottom}%
   \subcaptionbox{Tri-Linear}{\includegraphics[width=0.3\textwidth]{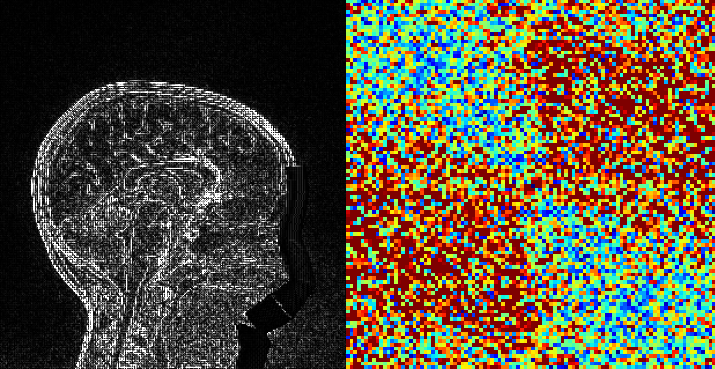}}
 \subcaptionbox{ESRGAN-3D}{\includegraphics[width=0.3\textwidth]{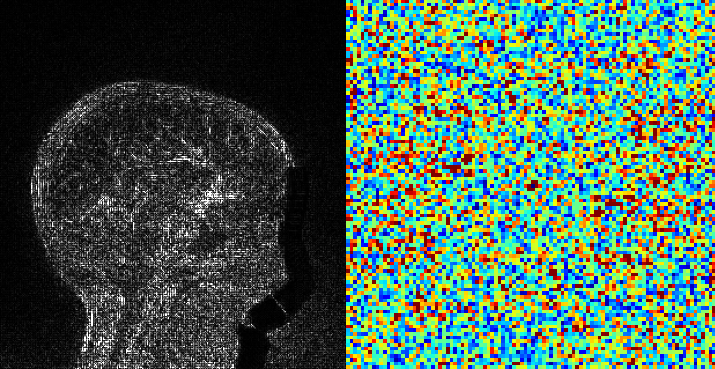}}
 \subcaptionbox{DCSRN}{\includegraphics[width=0.3\textwidth]{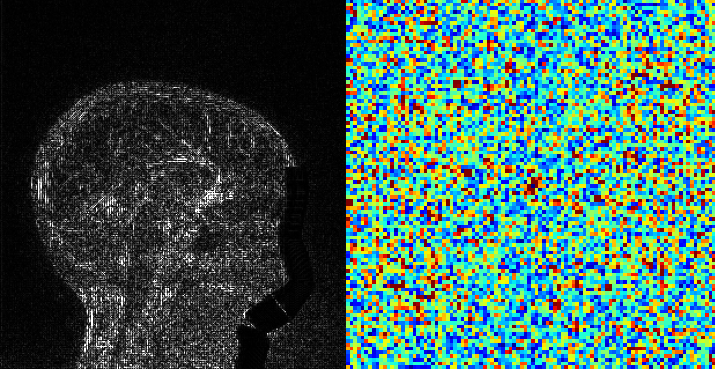}}
 \subcaptionbox{Wang et al.}{\includegraphics[width=0.3\textwidth]{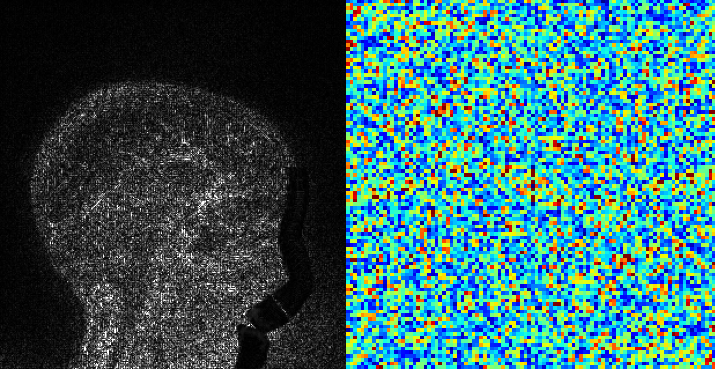}}
 \subcaptionbox{DISGAN (ours)}{\includegraphics[width=0.3\textwidth]{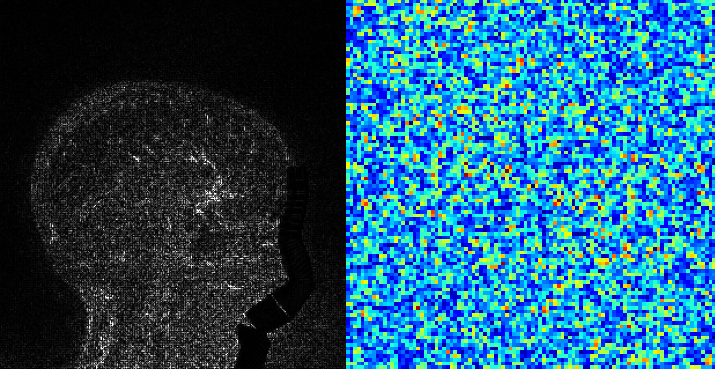}}
 \subcaptionbox{}{\includegraphics[width=0.3\textwidth, height=2.7cm,trim={1cm 0 0cm 13cm},clip]{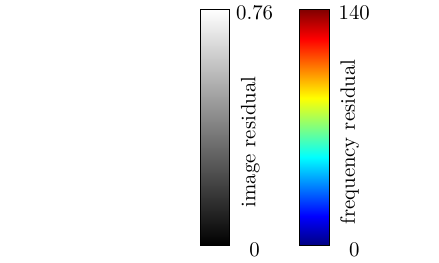}}%  
 \caption{Residual plots in both image and spatial frequency space of the same SR slices in Figure~\ref{fig:insample} over different methods. Each odd numbered column illustrates the absolute residual for SR methods against GT; each even numbered column illustrate the absolute residual of images in center-cropped frequency space of each models against the GT's. Note our model (DISGAN) shows the minimum residual in both image space and frequency space.}
 \label{fig:residual}
\end{figure*}
The results are shown in Figure~\ref{fig:insample}. Note that our model recovered the ground truth best, providing high quality anatomical structures with less noise. This is particularly true for the detailed blood vessel structures, which is reflected by the smallest residual in the zoomed region; our model also performs dominantly by recovering the finest structures of the cerebellum. The quantitative results are shown in Table~\ref{table:insample}. 

Furthermore, as shown in Figure~\ref{fig:residual}, the DWT-informed discriminator of our model guides the generator to produce images with high fidelity in both image space and frequency space, which is obtained by Fourier transforming the magnitude image without explicitly constraining the frequency space data. We crop the center matrix of $50\times 50\times 50$ of the frequency space of the image for better contrast when visualising the difference.

%-------------------------------------------------------------------------

\subsection{Removal of simulated noise}
In this experiment, we test the robustness of the model for the SR task on an image corrupted by simulated noise. The Gaussian noise is added to the HR image to simulate the noisy data, then the image is linearly downsampled by a factor of two as the LR input to the model. All models are dedicated SR models without any training on the denoising task, so the simulated noisy data can be considered as out-of-distribution data for the models. The models are evaluated on images added with four levels of increasing noise, whose standard deviation starts at $0$ and increases by $0.1$ per level to $0.3$.   

\begin{figure*}[h]
   \begin{center}\captionsetup[subfigure]{%singlelinecheck=false, 
   position=bottom}%
   \subcaptionbox{DISGAN (ours)}{\includegraphics[angle=270, width=0.24\textwidth]{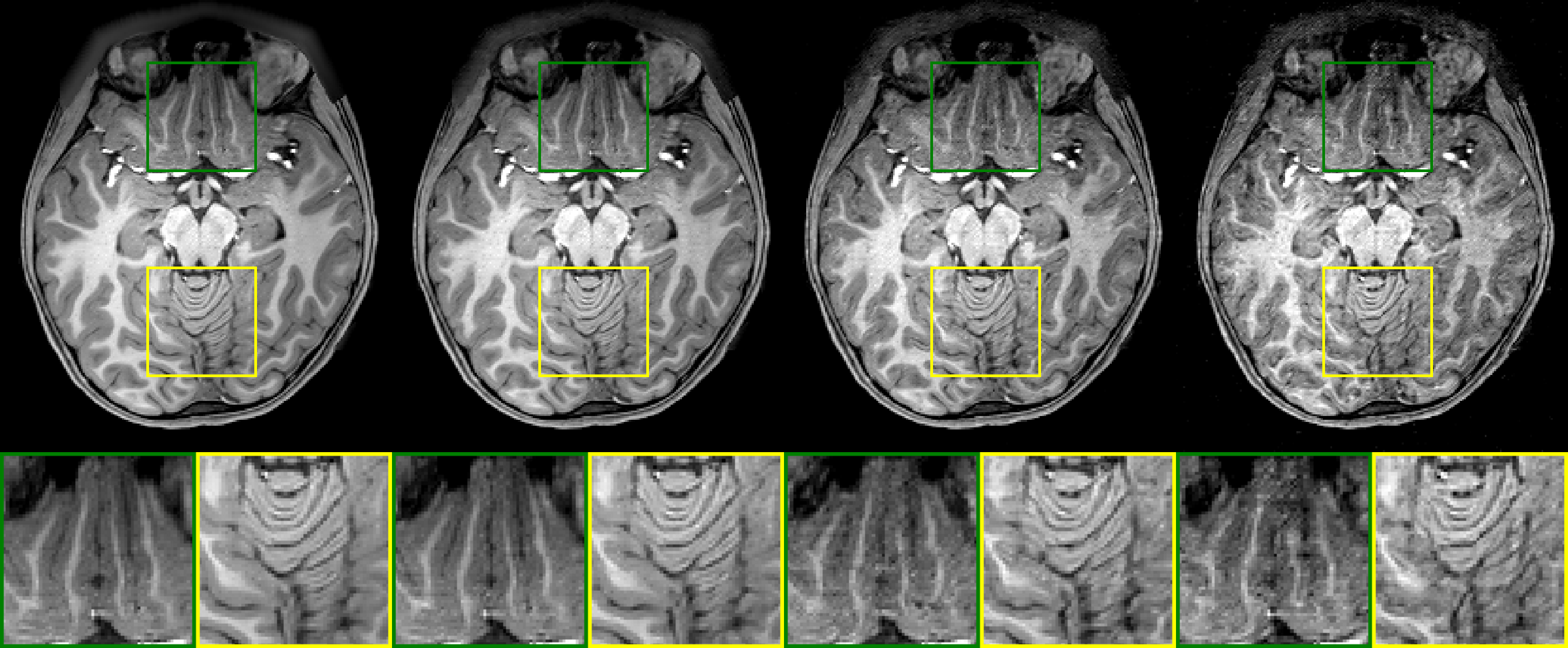}}
   \subcaptionbox{DCSRN}{\includegraphics[angle=270, width=0.24\textwidth]{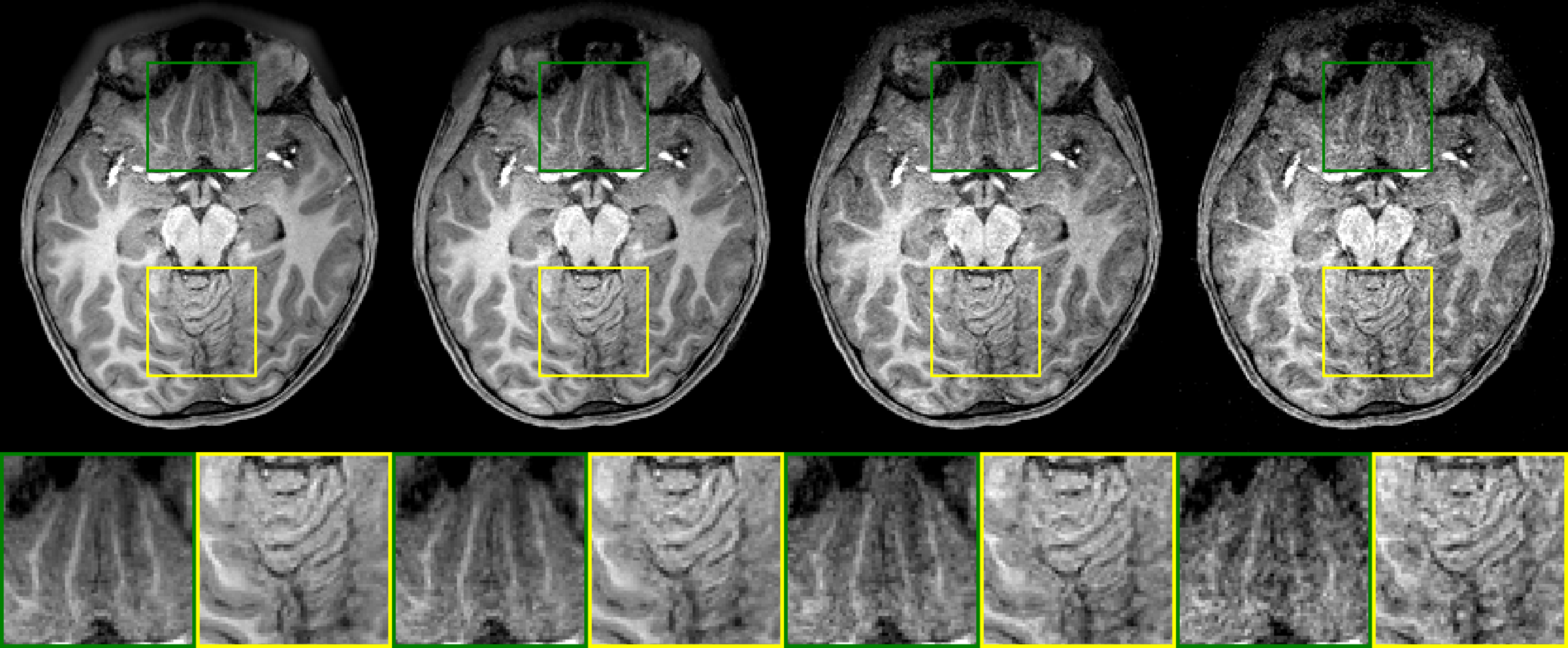}}
   \subcaptionbox{Wang et al.}{\includegraphics[angle=270, width=0.24\textwidth]{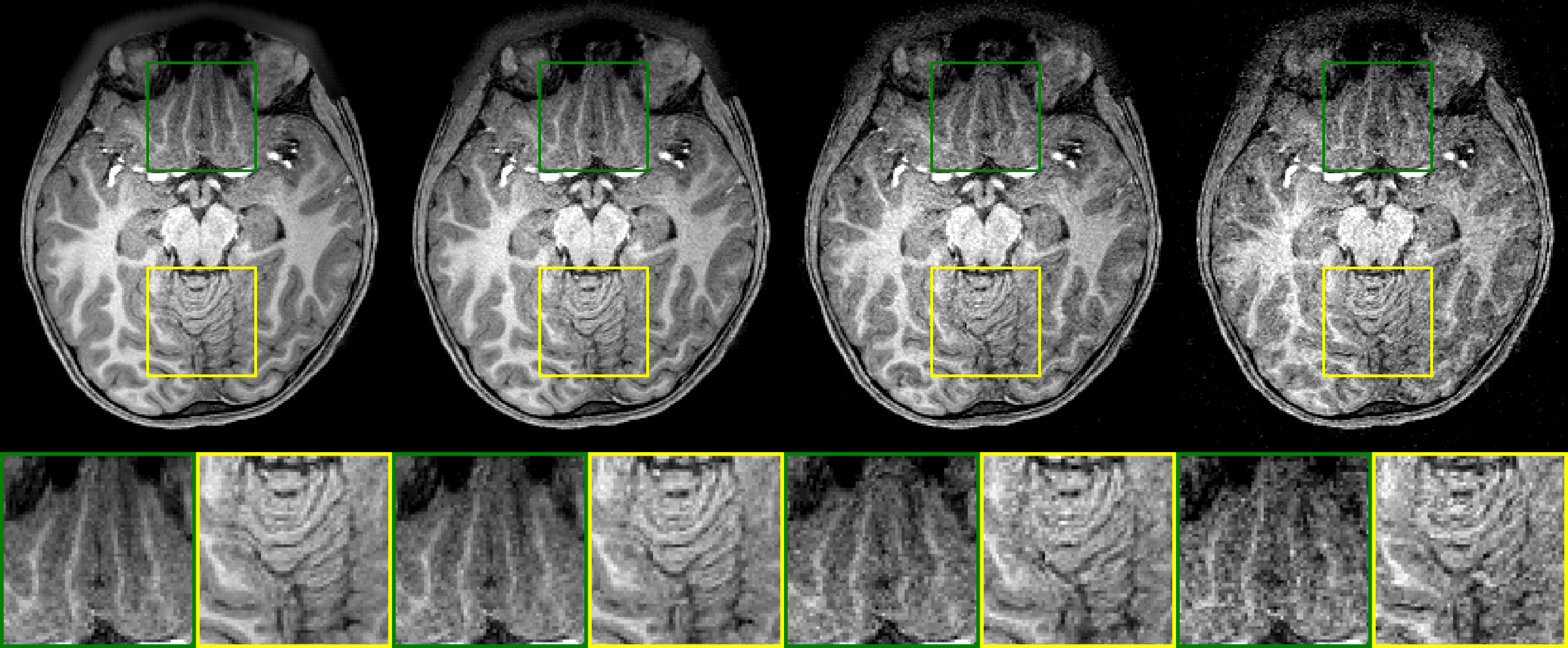}}
   \subcaptionbox{GT}{\includegraphics[angle=270, width=0.24\textwidth]{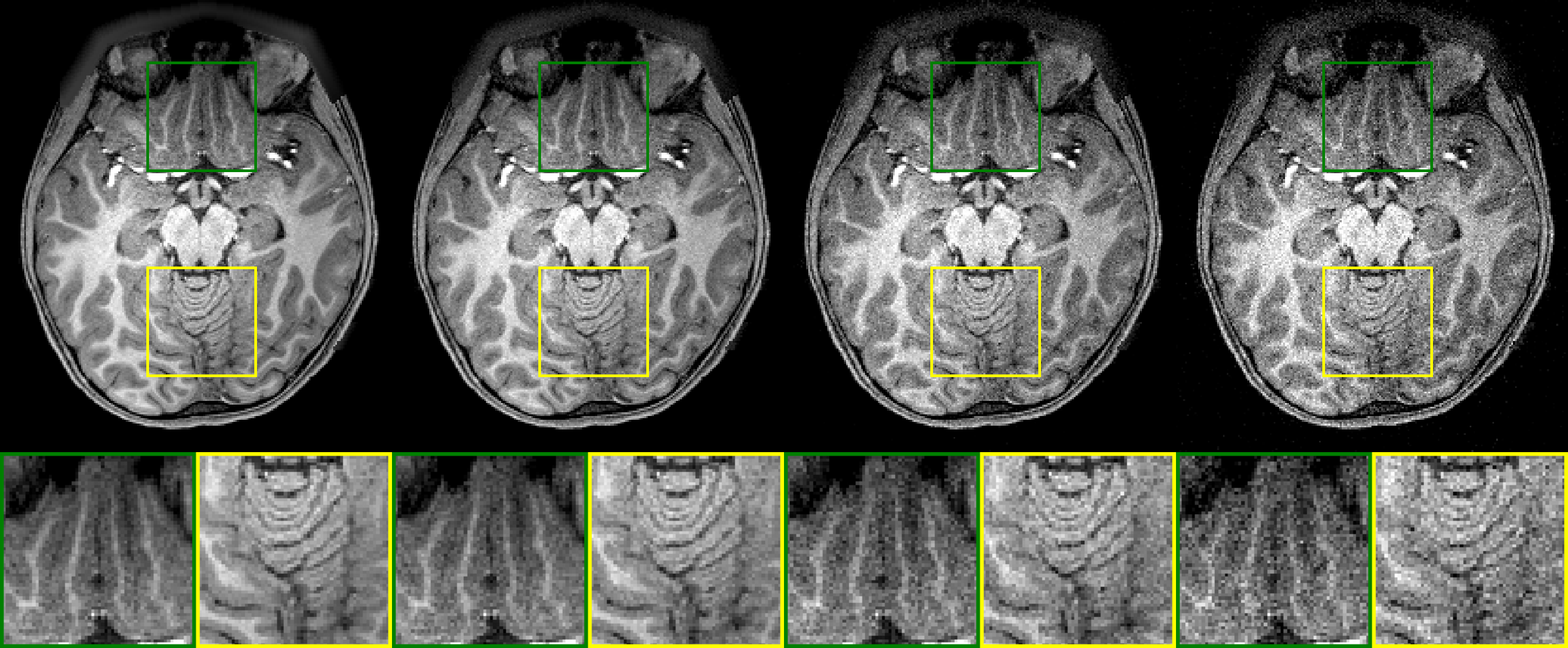}}

   \end{center}
      \caption{Comparison of different SR models on images with simulated noise. From \textit{left} to \textit{right}, each column stands for SR results from DISGAN (ours), DCSRN~\cite{DCSRN}, Wang et al.~\cite{ours_tmi}, and GT; from \textit{top} to \textit{bottom}, each row stands for images with different levels of noise with increasing standard deviation of $\sigma=0, 0.1, 0.2, 0.3$. Note the difference in zoom-in views of the region marked in yellow and green, with corresponding quantitative results shown in Figure~\ref{fig:metrics}. DISGAN (ours) keeps clean image contents throughout the noise-adding scheme.}
   \label{fig:sim_noise}
   \end{figure*}

\begin{figure}[h]
      \captionsetup[subfigure]{%singlelinecheck=false, 
      position=bottom, labelformat=empty}%
      \subcaptionbox{}{\includegraphics[height=0.21\textwidth]{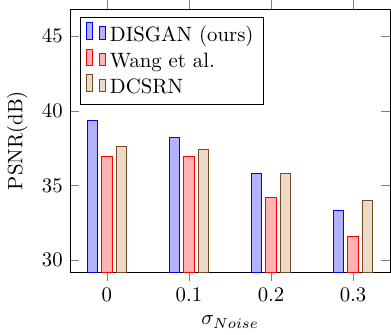}}%
      \subcaptionbox{}{\includegraphics[height=0.21\textwidth]{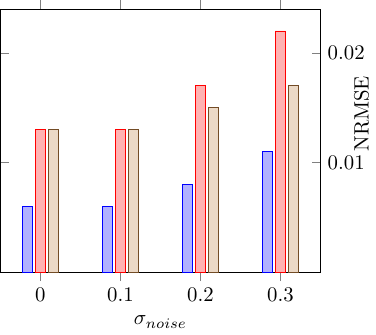}}
      \caption{The bar plots of PSNR (left) and NRMSE (right) metrics from different SR models on simulated noisy dataset, in accordance with Figure~\ref{fig:sim_noise}. For PSNR the higher the better, for NRMSE, the lower the better. DISGAN (Ours), Wang et al.~\cite{ours_tmi}, and DCSRN~\cite{DCSRN} are colored in blue, red, and yellow respectively.}%
      \label{fig:metrics}
   \end{figure}

As shown in Figure~\ref{fig:sim_noise}, our model produces SR images with clean content during the first three levels. At level 4, the performance of these models is most severely degraded, with all methods failing to maintain anatomical correctness (\eg some white matter structures are blurred, see region in yellow boxes of Figure~\ref{fig:sim_noise}). On the other hand, as shown in the metrics profile in Figure~\ref{fig:metrics}, our model outperforms other methods by a wide margin in the first two noise levels. It also maintains a low NRMSE value, reflecting a minimal residual against the GT, in line with its image quality.% 

   \begin{figure}[h]
      \begin{center}
         \subcaptionbox{GT}{\includegraphics[width=1\linewidth]{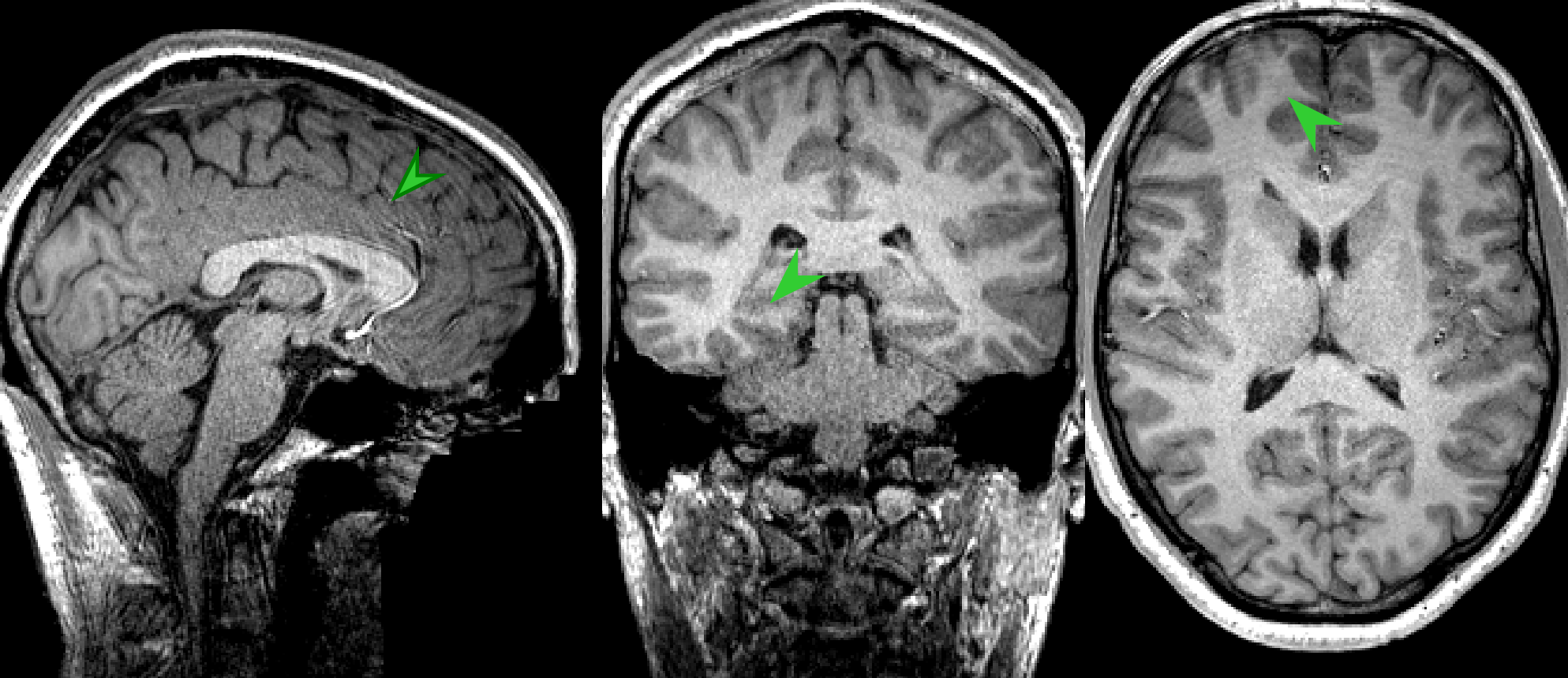}}
         \subcaptionbox{DISGAN (Ours)}{\includegraphics[width=1\linewidth]{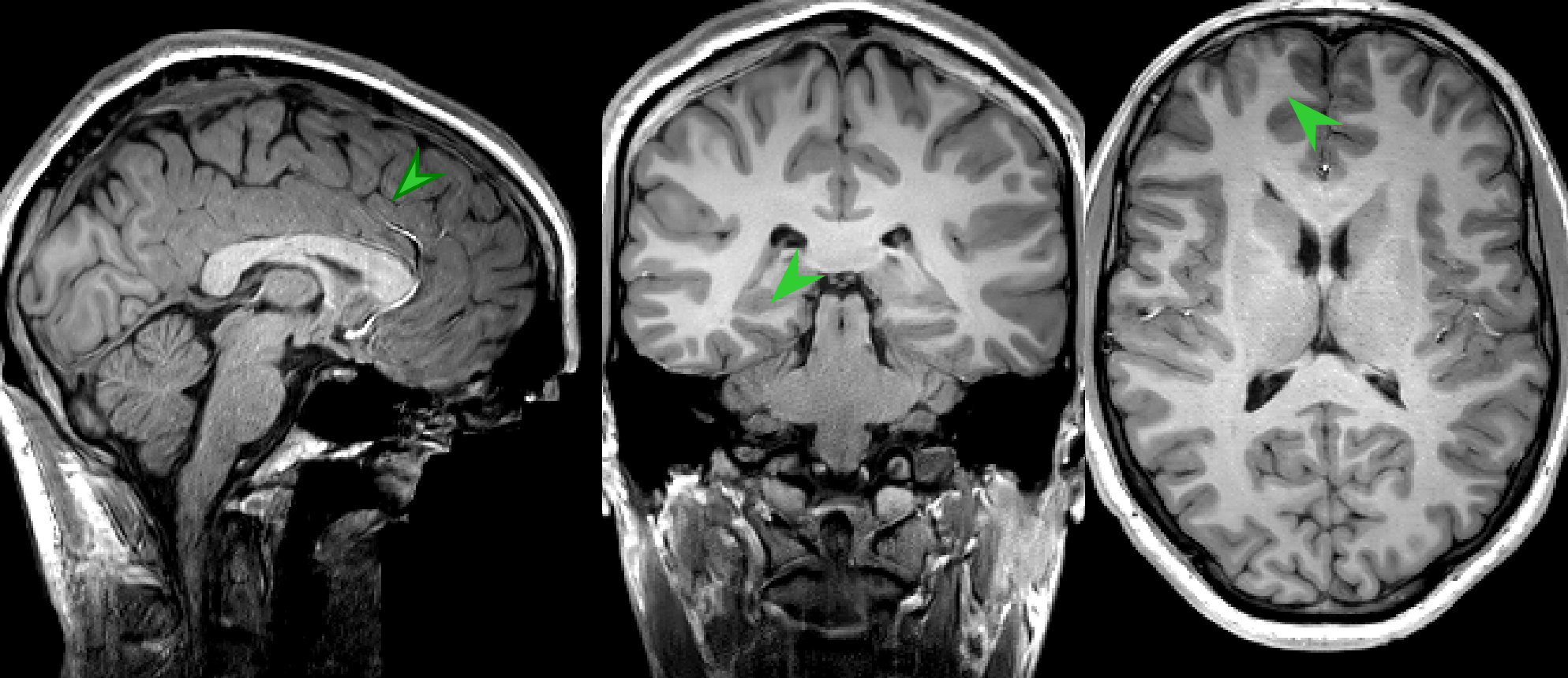}}
   
      \end{center}
         \caption{Noise cleaning of our DISGAN model on real world epilepsy noisy data. (a) the Ground truth image with random noise and ringing artefacts, indicated by the green arrows; (b) same anatomy after the noise removal by our DISGAN model. Note our DISGAN output images with more homogeneous intensity in gray and white matters, as well as minimal random noise and ringing artefacts.}
      \label{fig:epilepsy}
      \end{figure}
\subsection{Noise removal in real world patients}
In this experiment, we tested our model in recovering the anatomical details while performing noise cleaning on real-world patient MRI images, which are an unseen dataset for DISGAN and were acquired using different setups. We used MRIs of one subject each from the "Tumor" and "Epilepsy" datasets for the noise cleaning tasks. The generation process is identical to the previous experiments, which involves downsampling HR to an LR volume and generating SR from it. 

\textbf{Noise removal on "Epilepsy" dataset.}
As shown in Figure~\ref{fig:epilepsy}, the GT of the image contains random noise throughout and ringing artefacts in the prefrontal area of the cortex. Comparably, our DISGAN model reduces both random noise and ringing artefacts to a minimum, restoring a clean anatomical content. 

\begin{figure}[ht]
      \begin{center}
         \subcaptionbox{GT}{\includegraphics[width=1\linewidth]{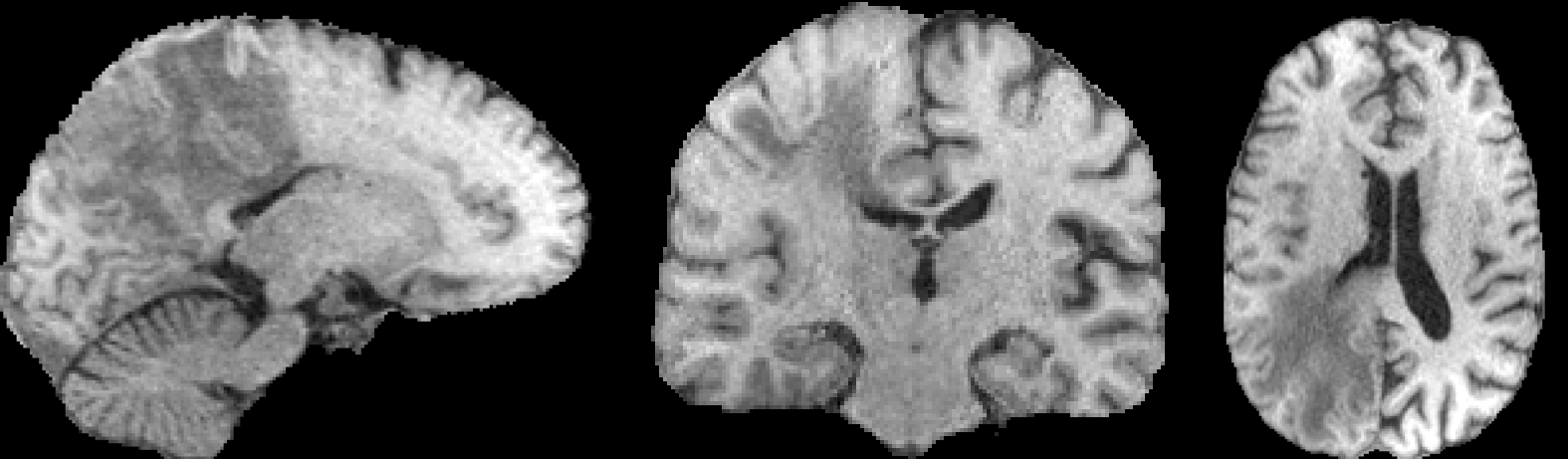}}
         \subcaptionbox{DISGAN (ours)}{\includegraphics[width=1\linewidth]{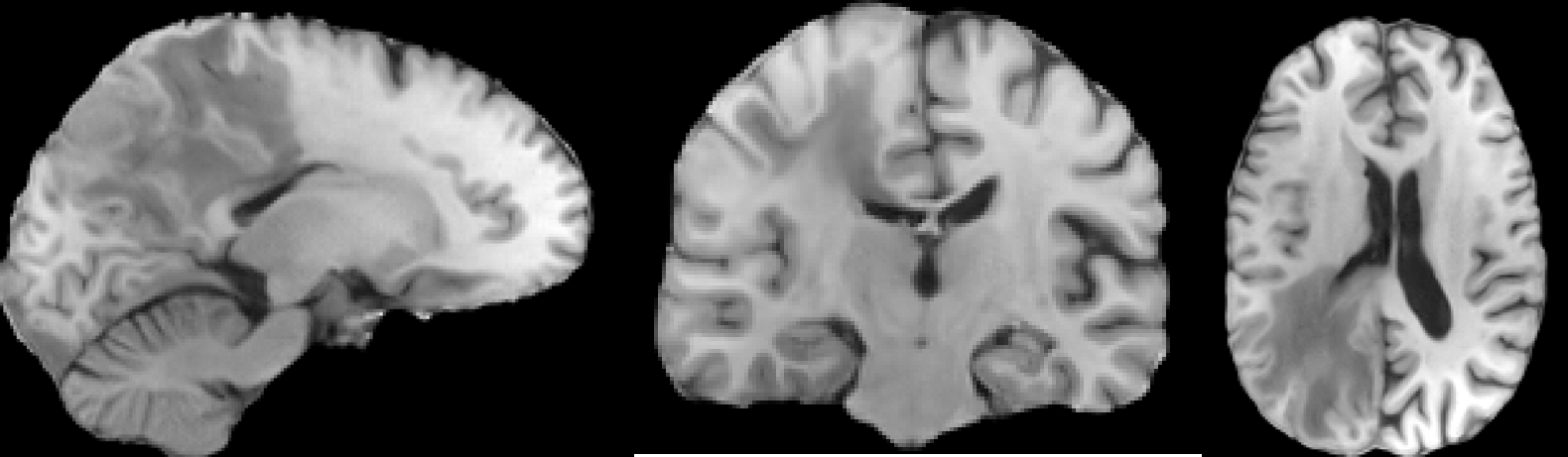}}
      \end{center}
      \caption{Noise cleaning of our DISGAN model on real world brain "Tumor" noisy data. (a) the Groud truth image with random noise degrades the image quality, corrupting the visibility of the structures; (b) the same anatomy after the noise cleaning by our DISGAN model.}
      \label{fig:brats}
      \end{figure}

\textbf{Noise removal on the "Tumor" dataset.}
In this experiment, we use a T1w brain MRI image from a subject in the "Tumor" dataset. Figure~\ref{fig:brats} shows the MRI with the patient's brain tumour and has noisy image content, which makes it difficult to identify the tumour. After the denoising SR process, our DISGAN model effectively restores the image content with minimal noise, representing a high quality anatomical structure of the brain. In addition, the contour of the tumour becomes more visible and the overall contrast for grey and white matter is improved, making visual inspection easier.
\section{Ablation studies}
For a detailed understanding of the effects of each component, we perform two ablation studies on the generator and discriminator building blocks. For the generator, we compare the generating ability of Volumetric RRDB (VRRDB) blocks, modified from its 2D RRDB block, and the 3D version of the SwinIR model. For the discriminator, we compare the effect of the DWT+conv unit in guiding high fidelity denoising SR generation.  
%-------------------------------------------------------------------------
\subsection{Building blocks in the generator}
To understand the architectural effects of the generator, we implemented and compared our generator with the SwinIR~\cite{SwinIR} generator for our SR task. SwinIR is the state-of-the-art model for 2D SR tasks, based on the efficient SwinTransformer blocks~\cite{SwinTransformer}. We reimplemented the SwinIR generator by replacing the 2D convolutions with 3D ones, and created tokens of size $4\times 4\times 4$. Due to the lack of 3D data pre-training, we trained the SwinIR generator from scratch on the same "Insample" dataset, on patched MRI volumes from the "Insample" dataset, with the remaining components of the GAN unchanged.

\begin{figure}
   \centering
   \captionsetup[subfigure]{position=bottom}%
   \subcaptionbox{GT}{\includegraphics[width=0.32\linewidth]{./trans_fig/GT_overlay.png}}
 \subcaptionbox{Ours+VRRDB}{\includegraphics[width=0.32\linewidth]{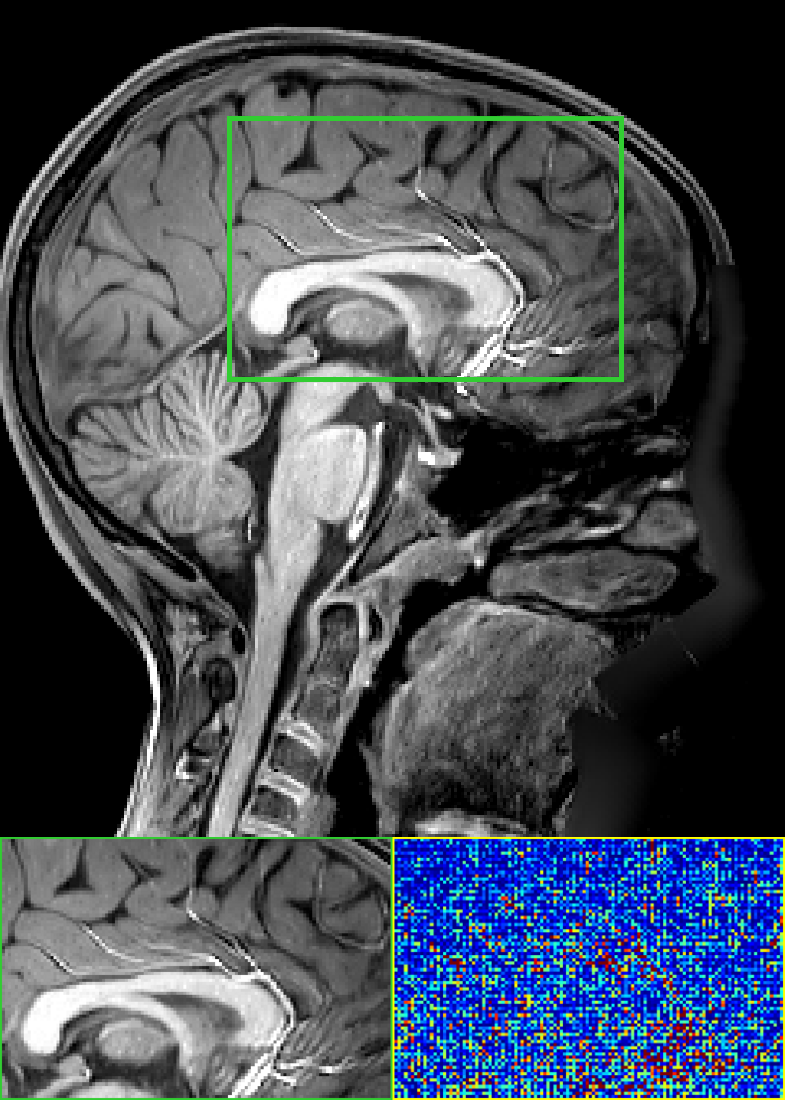}}
 \subcaptionbox{Ours+SwinIR}{\includegraphics[width=0.32\linewidth]{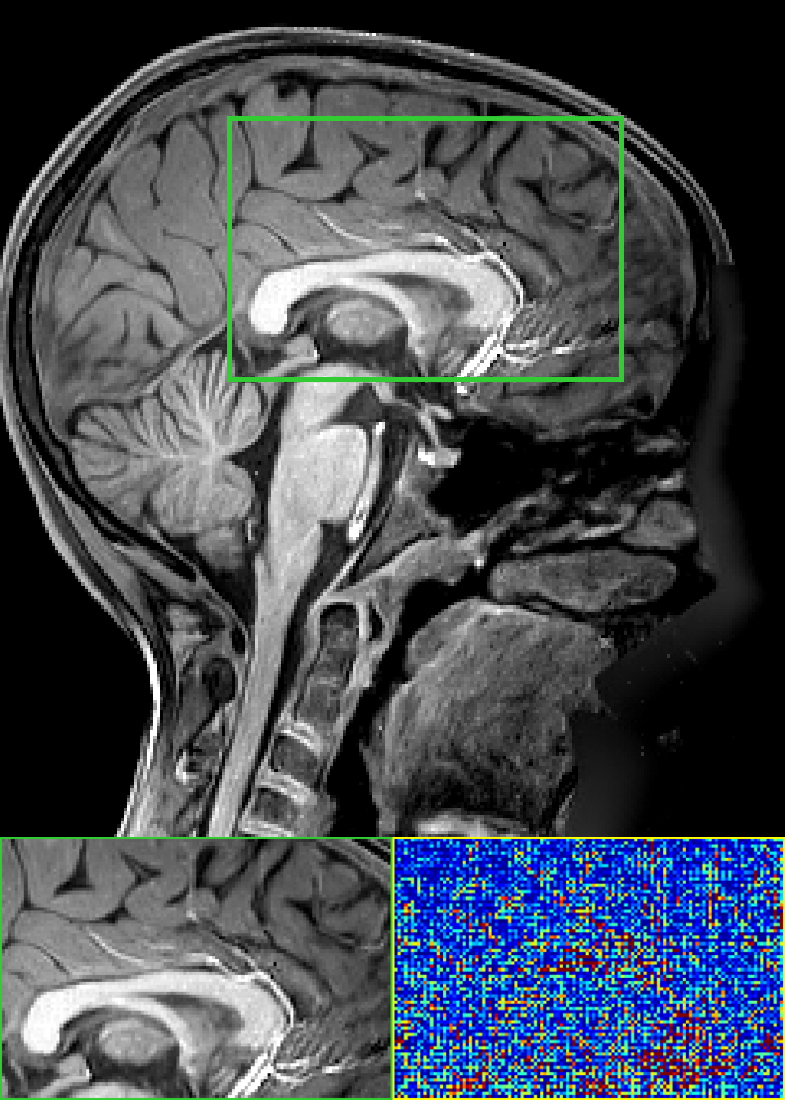}}%  
 \caption{Residual plots of SR results from SwinIR generator and VRRDB. The first row shows the sagittal view of the SR results, the second row shows zoom-in region and its absolute residuals.}
 \label{fig:abla_g}
\end{figure}

The results show a better recovery of details in the anatomical structures in the brain MRI. As shown in Figure~\ref{fig:abla_g}, our VRRDB generator outperforms the SwinIR generator by completing details in the tail of blood vessels (see zoom-in views and residual maps). We speculate that the reason for this is that the essence of the SR task lies in approximating local information rather than paying attention at a global scale.
%-------------------------------------------------------------------------
\subsection{DWT Unet as denoising discriminator}
In this section, we show the different images produced by models with different discriminators. These include a simple convolutional network, an Unet-shaped convolutional network~\cite{Unet}, and an Unet-shaped convolutional network with DWT+conv units. Our DISGAN model has the same generator architecture as Wang et al.~\cite{ours_tmi}, which has a normal multiple convolutional layers. As shown in Figure~\ref{fig:insample} and Figure~\ref{fig:sim_noise}, our model achieves cleaner image content and more structural details, reflecting the design capability of the DWT-informed discriminator.
%------------------------------------------------------------------------
\section{Discussion}
In this paper, we propose DISGAN, a GAN architecture for SISR and denoising of 3D MRI without the need for separate denoise training. Specifically, we propose an effective 3D DWT+conv block as a fundamental unit of our discriminator, which can indirectly guide the generator to output an image with high frequency fidelity and minimal noise. Our experimental results prove that our DISGAN model achieves distinguishable SR results with the closest detail restoration, while minimising noise. 

However, it is still an open question to understand the design factor leading to the generalisability of GAN models. In the future, we also plan to investigate the maneuverability of DISGAN for dedicated denoising tasks and improve the denoising ability for such unified model.

\subsection*{Acknowledgements}
This work is supported by the Deutsche Forschungsgemeinschaft (DFG - German Research Foundation), the grant numbers are DFG LO1728/2-1 and DFG HE 9297/1-1.  

{\small
\bibliographystyle{ieee_fullname}
\bibliography{egbib}
}

\end{document}